# Why The Trans Programmer?

Skye Kychenthal

skychenthal@gmail.com

March, 2022

## Abstract

Through online anecdotal evidence and online communities, there is an in-group idea of trans people (specifically trans-feminine individuals) disproportionately entering computer science education & fields. Existing data suggests this is a plausible trend, yet no research has been done into exactly why. As computer science education (traditional schooling or self-taught methods) is integral to working in computer science fields, a simple research survey was conducted to gather data on 138 trans people's experiences with computer science & computer science education. This article's purpose is to shed insight on the motivations for trans individuals choosing computer science paths, while acting as a basis and call to action for further research.

**Introduction**

Although not always the case [1], a binary gender gap in computer science currently exists with 80% of CS jobs filled by men. [2] As the majority of data exists within the gender binary; it is difficult to measure trans-specific data across the computer science gender gap. Still, with the little data available, it is shown trans individuals disproportionately enter some STEM fields. In the 2017-2018 curricular year, 6.7% of all U.S. degrees earned were in the Physical Sciences, Mathematics, or Computer Science [3][4][5], yet for those same degrees, the percentage of trans identifying individuals receiving them is 8.7%, and for specifically transgender individuals, the precentage is 10.9%. [6] StackOverflow's annual developer survey indicates 1.92% of their US developers are transgender, and another 1.42% are non-binary or gender nonconforming [7][B] compared to US states averaging roughly a 0.3% to 0.78% trans populus. [8] Beyond this, anecdotal evidence from online communities in spaces such as Reddit and Discord exist. Reddit is a social media board where users can share images, videos, and text in specific groups called subreddits. The subreddit r/transprogrammer is one such board for trans individuals interested in computer science with 8.3 thousand members as of March, 2022. Discord is a group-based social media platform housing a plethora of intersectional trans and programming communities. Though data and anecdotal evidence based on opinions & online group sizes exists, research into why trans individuals disproportionately enter computer science education and fields is scarce at best. Research on trans individual's motivations for entering computer science could provide better insight into the societal forces causing this phenomenon.

## Methods

A brief survey was created to explore this question. The survey would be disseminated throughout online groups such as Reddit and Discord; with targeted groups such as r/transprogrammer. Discord would be used as a supplement to target various trans programming communities. With under 1% of the United States population identifying as trans [8], these distribution methods attempt to compensate for the discrepancy in population and sample sizes between cisgender and trans people. Issues with these methods arise from the biases intrinsic to online communities such as Reddit & Discord: including demographics, interests, and levels of intrinsic motivation. An example of demographic biases stem from the age distribution of respondents, with the majority (57.2%) between the ages of 18 and 24 and 64.5% currently students. Biases of intrinsic motivation stem from 92.7% of participants stating they enjoy or greatly enjoy computer science and further biases lend themselves to the fact respondents responded to this survey. Further issues arise from the inherent differences between online and offline trans spaces. Memes such as "cat-girls", "programming socks", etc. are prevalent among online trans communities, but scarcely exist within offline spaces such as a university campus. To gauge the respondent's preference for online trans communities, the question "how experienced are you with catgirls?" was posed to which 80.5% of respondents were "kinda" or "very" experienced with cat-girls. Dissemination through online communities was necessary to achieve the largest sample size possible with the resources available.

## Results

Within the trans communities polled, there is a large discrepancy between those Assigned Male at Birth (AMAB) and Assigned Female at Birth (AFAB). Although Intersex was a polling option, no respondents identified as intersex. Assigned Gender at Birth (AGAB) was an optional question due to comfort reasons, still, a majority (81.2%) of respondents answered. Among those who answered, there was an overwhelming majority of respondents who identified as AMAB at 89% or 113 of the total responses. Within the self-identification section, 80.4% identified as trans-femine or a trans woman, 6.5% identified as trans-masculine or a trans man, while 34.8% identified as non-binary or gender-queer. Furthermore, 71.6% of trans individuals polled believe there is likely a disproportionate amount of trans people in computer science, with 16.4% believing the contrary.

| AGAB | Res # | % of Res |
| --- | --- | --- |
| AMAB | 100 | 89.3% |
| AFAB | 12 | 10.7% |

Fig 1. Number of respondents identifying as Assigned Male at Birth (AMAB) versus Assigned Female at Birth (AFAB). No individuals marked themselves as intersex.

| Identity | Res # | % of Res |
| --- | --- | --- |
| Femme | 111 | 80.4% |
| Masc | 9 | 6.5% |
| NB | 48 | 34.8% |

Fig 2. Number of respondents identifying as trans-feminine / trans women (femme), trans-masculine / trans men (masc), and non-binary or gender-queer (NB).

Among some of the free-response "why did you start programming" answers, a few trends presented themselves. The primary trends for why trans people entered into computer science were some form of enjoyment or desire to create a game or project. Lesser trends surfaced, such as parental influence, schooling, or financial security. Financial security often influenced individual's entering the industry, rather than their learning of computer science. Parental influence was most often indirect rather than direct; such as looking up to a parent in computer science.

| Reason for Learning | # of Res. |
|---|---|
| General Enjoyment | 36 |
| Game-development / Creation | 24½ |
| Parental Influence | 15½ |
| Financial Reasons | 12 |
| Schooling & Teachers | 8 |
| Boredom | 4 |

Fig 3. "Why did you start programming" response trends. Responses that fit two categories count as ½ for both categories. Most responses falling into two categories had to do with financials or schooling combined with enjoyment or creation.

Two insightful free response answers were:
1. "At the time, it felt like just following my interest and going with the flow. Since I discovered I'm trans, I now also see that it was partially about avoiding gender dysphoria; you don't need to perform gender for a computer."
2. "At first, it was because it [looked] interesting. But now, as a trans person, I feel like programming brings me more than just happiness. It brings me the security that other jobs don't. I don't want to be [starving] to death just because some hirees think that being trans means I'm incompetent."

In regards to parental influence, a separate question was used to gauge how heavily a respondent's parents influenced their entering into CS education. The scale measured 1 ("Not At All") to 5 ("Very Much So"). 67.4% answered 1 or 2, 8.7% answered 3, 14.5% answered 4, and only 9.4% answered 5 ("Very Much So").

Regarding computer science education, 64.5% of respondents were students and not in the work-force. 40.6% of respondents were self-taught, with 56.3% of self-taught individuals learning before the age of 13, and 36.8% learning between the ages of 13 and 17. Combining self-taught and traditionally taught students, 78.4% of all respondents had begun the study of computer science by the time they were 18.

| Education | # of Res. | % of Res. |
|---|---|---|
| Self-taught | 56 | 40.6% |
| Middle School | 23 | 16.7% |
| High School | 33 | 23.9% |
| College/Uni | 26 | 18.8% |

Fig 4. Distribution of the time at which respondents began learning computer science.

| Self-Taught Age | # of Res. | % of Res. |
|---|---|---|
| 0 - 13 | 49 | 56.3% |
| 13 - 17 | 32 | 36.8% |
| 18 - 24 | 2 | 2.3% |
| 25+ | 4 | 4.6% |

Fig 5. Distribution of the age at which self-taught respondents began learning computer science.

**Discussion**

Considering only 1 in 10 trans individuals polled were trans-masculine or non-binary AFABs, and given computer science as a predominantly male field, it can be assumed that the factors contributing to this male dominated field occur primarily corrolated to sex rather than gender. It's even more revealing as cis women enter computer science at a rate of 18% [2], a rate higher than trans-masculine or non-binrary AFABs. Those assigned female at birth are under-represented across the gender spectrum. Within this polling, a majority of trans people recognize this phenomenon. Considering there is recognition of this phenomenon within the trans community, there is shockingly little research into why. Hypothesis exist as to why cis females do not pursue computer science: primarily the West's history and over-emphasis of computer science as masculine [9] comparative to other parts of the world. [10] This overarching narrative leads to further ideas of individual's perceptions of computer science as a nerdy, isolated, basement-dwelling hacker's career. [11] However, few concrete examples exist as to why trans-feminine individuals will enter computer science while trans-masculine individuals might not. A hypothesis explored is the possible effects of parental influence. However, 76.1% of participants claimed parental influence did not have or had neutral bearing on their choice to pursue computer science. A more supported hypothesis as to why trans individuals enter computer science corralitive to sex rather than gender may not be to do with the societal influences themselves, but when those influences were introduced. Among millennial individuals, the average age of social transition is 20.77. And, although sooner for Gen Z individuals, the average age of social transition is 16. [12] Considering 78.4% of respondents begin learning before they are 18 and 39.5% learning roughly before middle-school, many trans individuals may not know they are trans or may be closeted when beginning to learn computer science. For a large portion of the respondents, society willl have acted towards these closeted or non-presenting individuals as their assigned gender when first introducing them to computer science. And considering adolescence is a critical period for gender socialization [13] the socialization of computer science as masculine and a nerdy male's job may have already occurred. This socialization is unfavorable to the prototype of a programmer, therefore, it stands to reason even a trans man would not want to follow this unfavorable hacker prototype of a male. Trans women who at the time present as males, will be treated and taught computer science with the socialized bias of a teenage boy. This could also be demonstrated by the respondents learning computer science for reasons of enjoyment as those learning later in life may be more inclined to learn CS for financial reasons. Combined with the ideas that computer science allows individuals the freedom to create distant from self-expression and identity, and with the eventual realization that computer science provides good financial security, all come together to make computer science a tempting career for many trans people.

**Conclusion**

Trans people are an integral part of society and also STEM. As a result of the fact trans-feminine individuals and AMABs are disproportionately represented in computer science compared to AFABs, more research needs to be done into why, and the implications this holds. This research may help address the societal issues of the gender gap in computer science with a broader and more detailed perspective, given issues of the gender gap exist across the gender spectrum. Because the teaching of computer science for many individuals begins at a young age, encouraging more AFAB individuals into computing, and the equalization of this male dominated field should begin with opportunities for AFABs at a young age. A trans-aware lens of analysis must be encouraged when conducting educational research, as trends within the trans community are significant and can shed light on issues within more gender binary focused research.

**Appendix A Additional Questions**

One of the unconventional questions from the survey was "what kind of games do you like?" This stems from a hypothesis that trans people or Redditors in general are more attracted to video and table-top role-playing games as well as strategy games. The results demonstrate 54.3% of trans individuals polled enjoyed role-playing video-games (RPGs), 42% enjoyed strategy games, and 55.8% liked what are termed "casual games." Only 8.7% of participants stated they do not play video games. Most trans people (55.4%) began playing video games aged 8 or younger, while another 35.4% began before 13.

**Appendix B StackOverflow Data**

The StackOverflow 2021 developer survey used polls for transgender and gender non conforming / non-binary developers across the globe. For the question "are you transgender?", only 1.28% of individuals responded yes, with another 0.75% responding "Or, in your own words…" A script was created to localize data to the United States in order to gather more comparable metrics. This script utalized the publicly available responses for the 80,000 user survey, filtered for the 20,000 US responses, and totalled the number of respondents reporteing "yes" to being transgender as well as reporting to be gender-non-conforming or non-binary. There could possibly be overlap between non-binary individuals & those who identified as transgender. Further, those who answered the "are you transgender?" question with "or, in your own words…" may have described themselves more specifically within the trans umbrella. The conservative estimates used in this paper still show a large trend, but the U.S. trans population on StackOverflow could be over 4% if including those who identified in their own words combined with minor overlap between non-binary and transgender identifications.

**Appendix C Resources**

StackOverflow Script:

  github.com/SkyMocha/StackOverflowData
  *The script for curtailing StackOverflow responses.*

r/transprogrammer Dataset:

  osf.io/5mh8u/
  *r/transprogrammer survey data in PDF format.*

# Citations

[1] Light, J. S. (1999). When Computers Were Women. *Technology and Culture*, *40*(3), 455–483. http://www.jstor.org/stable/25147356.

[2] "Women in Computer Science." ComputerScience.org, Feb. 2022.

[3] National Center for Education Statistics. Digest of Education Statistics, 2016. (n.d.). Retrieved from https://nces.ed.gov/programs/digest/d16/tables/dt16_322.50.asp

[4] National Center for Education Statistics. Digest of Education Statistics, 2019. (n.d.). Retrieved from https://nces.ed.gov/programs/digest/d19/tables/dt19_322.10.asp

[5] Published by Erin Duffin, & 23, F. (2021, February 23). U.S. higher education - number of bachelor's degrees 2030. Retrieved from https://www.statista.com/statistics/185157/number-of-bachelor-degrees-by-gender-since-1950/

[6] Greathouse, Maren & BrckaLorenz, Allison & Hoban, Mary & Huesman, Ronald & Rankin, Susan & Stolzenberg, Ellen Bara. Queer-spectrum and trans-spectrum student experiences in American higher education: the analyses of national survey findings. Retrieved from https://doi.org/doi:10.7282/t3-44fh-3b16

[7] "Stack Overflow Developer Survey 2021." *Stack Overflow*.

[8] Flores R. Andrew, et al. (2016) "How Many Adults Identify as Transgender in the United States?" *Williams Institute, UCLA*.

[9] Ensmenger N. (2015). "Beards, Sandals, and Other Signs of Rugged Individualism": Masculine Culture within the Computing Professions. *Osiris*, *30*, 38–65. https://doi.org/10.1086/682955

[10] Lagesen, V. A. (2008). A Cyberfeminist Utopia?: Perceptions of Gender and Computer Science among Malaysian Women Computer Science Students and Faculty. *Science, Technology, & Human Values*, *33*(1), 5–27. https://doi.org/10.1177/0162243907306192

[11] TURKLE, S. (1997). Computational Technologies and Images of the Self. *Social Research*, *64*(3), 1093–1111. http://www.jstor.org/stable/40971200

[12] Puckett, J. A., Tornello, S., Mustanski, B., & Newcomb, M. E. (2021). Gender variations, generational effects, and mental health of transgender people in relation to timing and status of gender identity milestones. *Psychology of Sexual Orientation and Gender Diversity*. Advance online publication. https://doi.org/10.1037/sgd0000391

[13] Balvin, N., Shivit Bakrania, S. O., & Kerry Albright, S. M. (2017, August 18). What is gender socialization and why does it matter?

# Acknowledgements

First, and foremost, I would like to thank the wonderful r/transprogrammer community for their interest and responses to my research. I would also like to thank John Reynolds (Summer Dragonfly) for disseminating the survey and providing feedback. I would like to thank u/UnrequitedMotivation for suggesting I look into the StackOverflow census. Finally, I would like to thank my teachers at The Peddie School as well as Kutub Gandhi for acting as spring-boards for ideas and rough drafts.